\documentclass[12pt]{article}
\usepackage{fullpage,amsmath,amssymb,titling,xcolor,graphicx,hyperref,cite}
\numberwithin{equation}{section}
\title{Defect entanglement entropy for superconformal RG Interfaces}
\author{}
\date{}

\newcommand{\dd}{\mathrm{d}}
\newcommand{\AdSLS}{AdS$_5^{\mathrm{LS}}$ }
\newcommand{\AdSN}{AdS$_5^{\mathcal{N}=4}$ }

\begin{document}

\begin{titlepage}
\vfill

\begin{flushright}
CCTP-2026-09\\
ITCP-2026-09
\end{flushright}

\begin{center}

\vskip .5in 
\noindent

{\Large\bf{Defect entanglement entropy for superconformal RG Interfaces}}

\bigskip\medskip

Evangelos Afxonidis$^{a,b}$\footnote{{\tt afxonidisevangelos@uniovi.es}}
and Christopher Rosen$^c$\footnote{{\tt rosen@physics.uoc.gr}}

\bigskip\medskip
{\small 

\textit{$^a$Department of Physics, University of Oviedo,
Avda. Calvo Sotelo s/n, 33007 Oviedo}}

\smallskip
{\small \textit{and}}

\smallskip
{\small 

\textit{Instituto Universitario de Ciencias y Tecnolog\'ias Espaciales de Asturias (ICTEA),\\
Calle de la Independencia 13, 33004 Oviedo, Spain}}

\bigskip
{\small 

\textit{$^c$Crete Center for Theoretical Physics, Department of Physics, University of Crete,\\
71003 Heraklion, Greece}}

\vskip 1cm 

     	{\bf Abstract }
	\end{center}
	We study the defect contribution to the entanglement entropy of a spherical region centred on four-dimensional superconformal RG interfaces. These interfaces are supersymmetric codimension one defects separating the Leigh--Strassler SCFT vacuum and the $\mathcal{N}=4$ SYM theory deformed by spatially varying mass terms. Exploiting the holographic dual to these interfaces, we extract a cutoff-independent interface contribution to the entanglement entropy, $\mathcal{C}^I$. We quantify the non-trivial dependence of $\mathcal{C}^I$  on the supersymmetric mass deformation, and observe a scaling regime for large deformations. We further relate $\mathcal{C}^I$ to the renormalized on-shell action and the stress-tensor one-point function of the interface theory. Our results provide the first fully backreacted holographic computation of this entanglement quantity for superconformal RG interfaces in more than two dimensions.
	\noindent

\noindent

\vfill
\eject

\end{titlepage}

\tableofcontents

\newpage

\section{Overview}
Deforming a quantum field through the introduction of a conformal defect offers a controllable means to relax a theory's dependence on spacetime symmetries. It is controllable in the sense that such systems may still possess a large amount of residual symmetry, which can be exploited to simplify calculations. In this way, one might hope to systematically depart from the Poincar\'e invariant setting, which is of both theoretical and phenomenological interest. In this work, we will exploit this perspective to investigate some properties of conformal defects of codimension one---interface conformal field theories.

If the ambient theory is a conformal field theory (CFT) in $d$ dimensions, insertion of a codimension one conformal interface preserves a $SO(d-1,2) \subset SO(d,2)$ subgroup of the ambient theory's conformal group. The resulting configuration is an example of a defect conformal field theory (dCFT). Perhaps the most familiar examples of such dCFTs are the Janus configurations of \cite{Bak:2003jk,Clark:2004sb,Clark:2005te,DHoker:2006vfr,Gaiotto:2008sd}, in which a conformal interface in the $\mathcal{N}=4$ SYM theory divides domains in which a marginal coupling (the gauge coupling/theta angle) assumes different values on either side of the interface. Depending on the details of the construction, such interfaces can also preserve supersymmetry.

It is also possible to construct conformal interface theories in which the ambient theories on either side of the interface are deformed by couplings for relevant operators. The spacetime dependence of such couplings are fully determined by the residual $SO(d-1,2)$ conformal symmetry. In particular, for a conformal interface located at $y=0$, a coupling $g(y)$ for a dimension $\Delta$ operator must have a profile of the form
\begin{equation}\label{eq:conDef}
g(y) = \frac{\lambda}{y^{d-\Delta}} \qquad \mathrm{with} \qquad \partial_\mu \lambda = 0
\end{equation}
on either side of the interface \cite{Herzog:2019bom,Herzog:2019rke,Bianchi:2019umv,Arav:2020obl}. 

In \cite{Arav:2020obl}, by applying the rigid conformal supergravity methods pioneered in \cite{Festuccia:2011ws}, it is demonstrated that such spatially dependent couplings for codimension one dCFTs are also compatible with supersymmetry. In particular, in \cite{Arav:2020obl} the conditions for a supersymmetric conformal interface in the $\mathcal{N}=4$ SYM theory are derived, in which the theory on either side of the interface is deformed by a spatially varying supersymmetric mass term. These couplings are of the form (\ref{eq:conDef}), with $\Delta =2, 3$ for the boson and fermion masses, respectively. The superconformal interfaces generically preserve $\mathcal{N}=1$ supersymmetry in three dimensions. In that work, holographic examples of such superconformal interfaces were constructed explicitly, by solving the BPS equations of the gravitational dual.

One interesting byproduct of these explicit holographic constructions was the discovery of a class of superconformal interfaces in which the interface separates two {\it distinct} ambient SCFTs. These configurations were subsequently studied in detail in \cite{Arav:2020asu}. Generically, they are superconformal interface configurations in which one side of the interface is the $\mathcal{N}=4$ SYM theory (possibly deformed by spatially dependent sources as above), while the other is the undeformed ``Leigh-Strassler'' (LS) $\mathcal{N}=1$ SCFT \cite{Leigh:1995ep}. Importantly,  the LS theory provides the IR endpoint of a Poincar\'e invariant renormalization group (RG) flow from a mass deformed $\mathcal{N}=4$ SYM in the UV. That the theories on either side of the interface enjoy this RG lineage in the Poincar\'e invariant setting distinguishes them---they are termed ``RG interfaces''. These privileged interfaces are introduced and discussed in further detail in \cite{Andrei:2018die,Brunner:2007ur,Gaiotto:2012np} as well as references therein.

The aim of this work is to uncover new features of holographic RG interfaces, extending the computations of \cite{Arav:2020asu} in several new directions. Specifically, we are interested in characterizing the interface's contribution to the entanglement entropy of a spherical entangling region centered on the interface. This quantity, which we call $\mathcal{C}^I$, has played an important role in the study of dCFTs. In particular, under certain assumptions, it has been shown to obey monotonicity along RG flows initiated by deformations localized on the interface's worldvolume (see e.g. \cite{Casini:2023kyj} for a helpful overview). In this sense, in such situations it is natural to ascribe to $\mathcal{C}^I$ an interpretation in terms of a number of ``degrees of freedom'' localized on the defect. 

The interface contribution to the sphere entanglement entropy has by now been studied in holographic dCFTs in myriad systems, largely facilitated by the ``user-friendly'' results of \cite{Jensen:2013lxa,Estes:2014hka}. While we will not attempt to comprehensively catalogue this extensive literature here; a broad survey is provided by \cite{Bachas:2001vj,Azeyanagi:2007qj,Hartman:2013mia,Gutperle:2015hcv,Jensen:2018rxu,Chen:2021mtn,Uhlemann:2023oea,Karch:2023evr,Giombi:2024qbm,Jokela:2025qac,Melby-Thompson:2017aip,Erdmenger:2015spo}. These works are unified through their study of the computation and features of the interface entropy in various dimensions, as well as its behavior along dCFT RG flow.  Our work is the first calculation of this entanglement contribution in a holographic RG interface in $d>2$ dual to a fully backreacted solution of type IIB supergravity. In adopting this perspective, we will confront new computational challenges---both technical and conceptual.

Taking cues from \cite{Casini:2023kyj, Kobayashi:2018lil,Chalabi:2021jud,Conti:2025wwf}, we further attempt to find non-trivial relationships between $\mathcal{C}^I$ and other field theory quantities. Performing a careful application of holographic renormalization \cite{deHaro:2000vlm,Bianchi:2001kw}, we demonstrate that the dCFT's renormalized on-shell action can be naturally written in terms of $\mathcal{C}^I$ and a weight $a_T$ characterizing the one-point function of the dCFT stress tensor (when the theory is placed on an AdS$_4$ background),
\begin{equation}
\langle \mathcal{T}_{ab} \rangle\,\dd x^a \dd x^b = a_T \,\dd s^2\left(\mathrm{AdS}_4 \right).
\end{equation}
This observation can be generalized straightforwardly for holographic dCFTs of other dimension and codimension. 

The plan for the rest of the paper is as follows: in section \ref{sec:RGIsols} we review the family of holographic RG interfaces introduced in \cite{Arav:2020asu}.  Specifically, we orient them within a consistent truncation of the maximal $SO(6)$ gauged supergravity in five dimensions, and describe the numerical recipe employed to construct the solutions. Exploiting their pedigree as solutions through oxidation to type IIB supergravity in ten dimensions, we detail the holographic interpretation of the bulk solutions from the perspective of the dual field theories.

In section \ref{sec:eCfun}, we give a self-contained recap of the techniques developed in \cite{Jensen:2013lxa,Estes:2014hka} for the holographic computation of sphere entanglement entropy in dCFTs. We then follow in section \ref{sec:4d} with an application of these methods to our family of RG interfaces. In doing so, we outline our algorithm for handling the short distance divergences inherent to the entanglement entropy. This task is complicated by the fact that our bulk solutions are known only numerically, and subtraction errors must be handled carefully. Our technique for dealing with these practical complications involves using the near boundary analytic form of the solution to perform the subtraction to high accuracy. We then summarize our results for $\mathcal{C}^I$ graphically, and study its asymptotic behavior, finding scaling regimes near the boundaries of the RG interface family. 

Section \ref{sec:obs} continues our holographic investigation into the physics of RG interfaces, with computations of both the stress tensor one-point function and the on-shell action for these dCFTs. Emphasis is placed on the relationship between these quantities and the defect contribution to the entanglement entropy, $\mathcal{C}^I$. We end the paper with a discussion in \ref{sec:dis}.

Supplementary material is housed in a series of appendices. In appendix \ref{app:FG}, exact expressions for the analytic form of the near boundary expansion of our solutions are provided. Appendix \ref{app:OS} contains the details of the holographic renormalization necessary for computing dCFT observables. Appendix \ref{app:RGscaling} investigates the behavior of $\mathcal{C}^I$ for mass deformations of large magnitude from the perspective of the details of the bulk geometry. Finally, in appendix \ref{app:schemes} we comment on the non-uniqueness inherent to the definition of the interface contribution to the entanglement entropy for these RG interfaces. We provide several alternative schemes, commenting on the qualitative features of $\mathcal{C}^I$ that appear to be robust to this inherent ambiguity.

\section{Holographic RG Interfaces}\label{sec:RGIsols}
In this section we quickly review the one-parameter family of holographic superconformal interfaces constructed in \cite{Arav:2020asu}. These solutions and their holographic interpretation are discussed in considerable detail in \cite{Arav:2020obl,Arav:2020asu}.
\subsection{The model}
The bulk solutions can be efficiently constructed within a five dimensional theory \cite{Bobev:2016nua} that arises as a consistent truncation of the $\mathcal{N}=8$ $SO(6)$ gauged supergravity theory. Working in a mostly minus signature for the metric, the gravity-scalar sector of this truncated theory can be described by the Lagrangian
\begin{equation}\label{eq:L5D}
e^{-1}\mathcal{L} = -\frac{1}{4}R+3(\partial\beta)^2+\frac{1}{2}\mathcal{K}_{z\bar{z}}\partial_\mu z \partial^\mu \bar{z}-\mathcal{P}.
\end{equation}
Here $\beta$ is a real scalar field, while $z$ is complex. The K\"ahler metric $\mathcal{K}_{z\bar{z}}$ descends from the K\"ahler potential in the usual way, 
\begin{equation}
\mathcal{K}_{z\bar{z}} = \partial_z\partial_{\bar{z}}\mathcal{K} \qquad \mathrm{where}\qquad \mathcal{K}=-4\log\left(1-z\bar{z} \right).
\end{equation}
The scalar potential $\mathcal{P}$ can be written in terms of a superpotential $\mathcal{W}$ like
\begin{equation}
\mathcal{P} = \frac{1}{8}e^\mathcal{K}\left(\frac{1}{6}\partial_\beta\mathcal{W}\partial_\beta\overline{\mathcal{W}} +\mathcal{K}^{z\bar{z}}\nabla_z\mathcal{W}\nabla_{\bar{z}}\overline{\mathcal{W}}-\frac{8}{3}\mathcal{W}\overline{\mathcal{W}}\right)
\end{equation}
where
\begin{equation}
L\,\mathcal{W} = e^{-4\beta}\left(1+6z^2+z^4 \right)+2e^{2\beta}\left(1-z^2 \right)^2.
\end{equation}

This theory admits a maximally AdS$_5$ vacuum with $\beta = z = 0$ and radius $L$. We will henceforth refer to this solution as \AdSN as it is the vacuum dual to the $\mathcal{N}=4$ SYM theory in its holographic limit. It further admits two AdS$_5$ vacua with $\beta = -\log(2)/6$ and $z = \pm i (2-\sqrt{3})$. These two solutions have AdS radius $\tilde{L} = 3 L/2^{5/3}$, and are related to one another by a $\mathbb{Z}_2$ symmetry of the bulk theory. These solutions are (both) dual to the Leigh-Strassler (LS) SCFT, and we will accordingly refer to them as \AdSLS with a $\pm$ subscript if there is an ambiguity in context.

For holographic purposes, it is helpful to parametrise the complex scalar $z$ in terms of two real scalar fields $\alpha$ and $\phi$, like
\begin{equation}
z = \tanh\left[\frac{1}{2}\left(\alpha-i\phi\right) \right].
\end{equation}
Fluctuations of these real scalar fields carry well defined quantum numbers under the symmetry group of the \AdSN solution uplifted on the five sphere to IIB, and are thus convenient for matching to scalar operators $\mathcal{O}^\Delta$ in the spectrum of the $\mathcal{N}=4$ theory. In particular, adopting an $\mathcal{N}=1$ organization of the $\mathcal{N}=4$ SYM field content, the dictionary is given by
\begin{align}
\phi \quad& \leftrightarrow   \quad\mathcal{O}^{\Delta = 3}_\phi  = \mathrm{tr}\big(\chi_3\chi_3 + \ldots \big) + \mathrm{h.c.} \nonumber \\
\alpha \quad& \leftrightarrow \quad \mathcal{O}^{\Delta = 2}_\alpha  = \mathrm{tr}\big(Z_3Z_3\big) + \mathrm{h.c.}\nonumber\\
\beta \quad& \leftrightarrow \quad  \mathcal{O}^{\Delta = 2}_\beta  = \mathrm{tr}\big(|Z_1|^2+|Z_2|^2-2|Z_3|^2 \big) + \mathrm{h.c.}
\end{align}
where $Z_a$ and $\chi_a$ are (respectively) the bosonic and fermionic components of the chiral superfield $\Phi_a$.
\subsection{The solutions}\label{sec:sols}
We will be interested in supersymmetric solutions to the theory described by (\ref{eq:L5D}) which manifest the preserved symmetries of the superconformal interfaces in their isometry group. This requirement motivates a bulk ansatz of the form
\begin{equation}\label{eq:anstz}
\dd s^2_5 = e^{2A}\dd s^2 \left(AdS_4 \right)-\dd r^2, \qquad \mathrm{and}\qquad \beta = \beta(r), \quad z = z(r)
\end{equation}
where the metric function $A$ depends only on the radial coordinate $r$. The metric on the leaves of the AdS$_4$ foliation can be written in Poincar\'e coordinates like
\begin{equation}\label{eq:AdS4g}
\dd s^2 \left( AdS_4\right) = \frac{\ell^2}{x^2}\left(\dd t^2 -\dd x^2- \dd r_{||}^2-r_{||}^2\dd \varphi^2 \right).
\end{equation}

Following \cite{Arav:2020asu}, we numerically construct a one parameter family of holographic RG interfaces by implementing the following algorithm:
\begin{enumerate}
\item Compute linearised fluctuations of the  \AdSLS fixed point which satisfy the BPS equations.
\item From these fluctuations, identify a mode which grows as one moves ``away'' from the \AdSLS fixed point along the radial direction.
\item Integrate this growing mode away from the \AdSLS by using it to seed a numerical shooting routine.
\end{enumerate}
The output of step 2 yields a BPS mode of the form
\begin{align}\label{eq:BPSm}
z &=i(2-\sqrt{3}) + i \zeta e^{r(1+\sqrt{7})/\tilde{L}}+\ldots \nonumber\\
\beta & = -\frac{1}{6}\log 2 + b\, \zeta e^{r(1+\sqrt{7})/\tilde{L}}+\ldots
\end{align}
where
\begin{equation}
b = \frac{1}{18}\left(3+2\sqrt{3} \right)\left(1+\sqrt{7} \right)
\end{equation}
and we have assumed that the \AdSLS boundary is obtained as $r\to-\infty$. Here $\zeta$ is a real parameter governing the amplitude of the linearised mode. Performing the integration in step 3 then leads to a one parameter family of supersymmetric solutions governed by the choice of $\zeta$. 

\begin{figure}[h]
\centering
\includegraphics[scale=1]{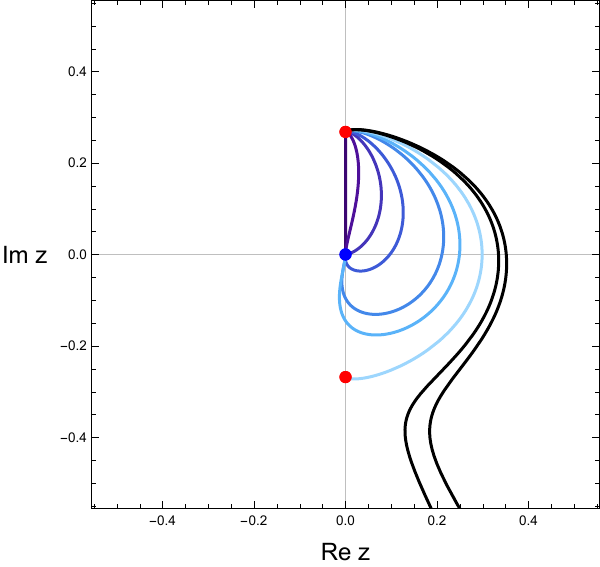}
\caption{Representative solutions spanning the solution space of the RG interfaces studied here (colored), together with a couple singular solutions (black) which lie outside our study. The solid red dots mark the location of the LS$^\pm$ vacua, while the solid blue dot corresponds to the $\mathcal{N}=4$ vacuum.}\label{fig:ImzvsRez}
\end{figure}

Some members of this one parameter family of solutions are illustrated in figure \ref{fig:ImzvsRez}. By construction, each solution in this family has (at least one) asymptotically \AdSLS boundary---the topmost red filled circle. After integrating the BPS mode (\ref{eq:BPSm}), one either arrives at an asymptotically \AdSN boundary, an \AdSLS boundary, or a singularity in the bulk. Our focus will be primarily on the first of these three possibilities---such solutions holographically describe an RG interface between the $\mathcal{N}=4$ SYM theory and the LS SCFT. The second possibility realizes a ``Janus'' type configuration, holographically dual to a superconformal interface in the LS SCFT. In the bulk, the solution asymptotes on either side to an LS boundary. The \AdSLS vacua obtained in the limit are related to one another by a bulk $\mathbb{Z}_2$ symmetry. The third possibility, in which the solution arrives at a bulk singularity somewhere along the radial direction, may have a holographic interpretation as a boundary CFT \cite{Gutperle:2012hy}, but will play no role in the present work.

\subsection{The holographic interpretation}
Given any holographic interface solution in this family, it is straightforward to perform a standard application of holographic renormalization (see e.g. \cite{Arav:2020obl}) to compute the sources and one-point functions that characterize the dual physical system. Here we will collect some useful results from this exercise, again referring the reader to \cite{Arav:2020obl, Arav:2020asu} for details. 

For illustrative purposes, we will choose coordinates such that an \AdSLS boundary is again obtained as $r\to -\infty$. In this case, one finds that the parameter $\zeta$ in (\ref{eq:BPSm}) governs the vacuum expectation values of a relevant scalar operator $\mathcal{O}^{\Delta = 1+\sqrt{7}}$, as well as that of two irrelevant scalar operators $\mathcal{O}^{\Delta = 2+\sqrt{7}}$ and $\mathcal{O}^{\Delta = 3+\sqrt{7}}$. Importantly, the BPS equations can be used to show that for these supersymmetric interface configurations no source can be turned on for any of these operators in the LS theory.

When the bulk solution has an asymptotically \AdSN boundary, the holographic interpretation becomes a bit more involved. If this boundary is obtained as $r\to\infty$, then any solution to the BPS equations will generically have the limiting behavior
\begin{align}\label{eq:Nis4asym}
A & = \frac{r}{L}  + \ldots \nonumber\\
\phi & = \phi_{(s)} e^{-r/L} + \ldots + \phi_{(v)} e^{-3r/L}+\ldots \nonumber\\
\alpha & = \alpha_{(s)}\frac{r}{L} e^{-2r/L} + \alpha_{(v)}e^{-2r/L}+\ldots \nonumber\\
\beta & = \beta_{(s)}\frac{r}{L} e^{-2r/L} + \beta_{(v)}e^{-2r/L}+\ldots 
\end{align}
where $\phi_{(s,v)}, \alpha_{(s,v)}, \beta_{(s,v)}$ are constants. The BPS equations imply relationships between these constants---they are not all independent in a supersymmetric configuration. Moreover, all terms included in the ellipses are vanishing in the $r\to\infty$ limit. In particular, we have employed a residual shift symmetry of the radial coordinate $r$ to remove any constant term in the asymptotic expansion of $A$. 

Solutions in the ansatz (\ref{eq:anstz}) are naturally associated with conformal boundaries which inherit an AdS$_4$ metric. Holographically, from this perspective they describe the physics of the dual SCFTs placed on an AdS$_4$ background. Here we are primarily interested in interfaces between SCFTs in flat space. We can thus perform a boundary Weyl transformation on the SCFT sources and vevs to obtain results relevant for the flat space interface. In practice, this manoeuvre is complicated somewhat by the presence of a conformal anomaly in the boundary theory. 

The upshot of a careful analysis \cite{Arav:2020obl} is that solutions asymptoting to the \AdSN boundary as in (\ref{eq:Nis4asym}) describe the $\mathcal{N}=4$ SYM theory deformed by the spatially varying sources\footnote{Here we work in conventions in which the Killing spinor projection paramater of \cite{Arav:2020obl} satisfies $\kappa = -1$.}
\begin{equation}\label{eq:holoSrc}
\frac{\ell \phi_{(s)}}{y_3}, \quad \frac{\ell^2\alpha_{(s)}}{y_3^2}, \quad \frac{\ell^2\beta_{(s)}}{y_3^2}\qquad \mathrm{with} \qquad \alpha_{(s)} = \frac{L}{\ell}\phi_{(s)}, \qquad \beta_{(s)} = -\frac{2}{3}\phi^2_{(s)}
\end{equation}
and where $y_3$, the direction normal to the interface, takes values $y_3 >0$. Evidently, all supersymmetric deformations in this class of solutions are controlled by the constant $\phi_{(s)}$. It will thus often prove convenient to parametrize our family of RG interface solutions in terms of the source $\phi_{(s)}$. A similar analysis leads to expressions for the one point functions of the field theory operators $\mathcal{O}_\phi^{\Delta = 3}$ and $\mathcal{O}_\alpha^{\Delta = 2}$, $\mathcal{O}_\beta^{\Delta = 2}$ to which these sources couple. 

\section{Interface entanglement entropy}\label{sec:eCfun}
In this section we briefly review expectations for the superconformal interface entanglement entropy, before describing the method we will use to compute it in the family of RG interfaces introduced above. 
\subsection{General considerations}
For a spherical entangling region of radius $R$ centered on the interface, the expected form of the entanglement entropy is in general \cite{Estes:2014hka}
\begin{equation}\label{eq:SEEgen}
S_{EE} \approx s_2 \frac{R^2}{\epsilon^2} + s_1 \frac{R}{\epsilon} + s_L \log\frac{2R}{\epsilon}+s_0 + \ldots
\end{equation}
where the ellipses denote terms that vanish as the UV cut-off $\epsilon$ is taken to zero. As explained in \cite{Jensen:2013lxa}, one can extract from this entanglement entropy a ``universal'' quantity $\mathcal{C}^I$ that characterises the contribution from the interface. To do so, one first subtracts from (\ref{eq:SEEgen}) the (half-space) contributions from the left and right sides of the interface, which we call $S_{EE}^{LS}$ and $S_{EE}^{\mathcal{N}=4}$ respectively\footnote{Note that for generic RG interfaces in our family, the $\mathcal{N}=4$ side is deformed by the supersymmetric bosonic and fermionic mass terms dual to the fall-offs in (\ref{eq:holoSrc}). Accordingly, as shown explicitly below, $S_{EE}^{\mathcal{N}=4}$ does not generically evaluate to the vacuum result.}:
\begin{equation}\label{eq:SEEI}
S_{EE}^I \equiv S_{EE}-\frac{1}{2}\left(S_{EE}^{\mathcal{N}=4}+S_{EE}^{LS} \right) \approx s_1 \frac{R}{\epsilon} + \mathcal{C}^I + \ldots
\end{equation}
The quantity $\mathcal{C}^I$ is universal in the sense that once this subtraction has been performed, rescaling the UV cut-off $\epsilon$ does not modify its value. This is to be contrasted with the constant $s_1$, which is {\it non}-universal. Our primary aim is to compute $\mathcal{C}^I$ for the family of holographic superconformal RG interfaces described in section \ref{sec:sols}.

\subsection{Defect entanglement entropy for holographic interfaces}
Holographic entanglement entropies for static spacetimes are computed via the area $\mathbb{A}_{\mathrm{min}}$ of the bulk minimal area surface which is affixed to the boundary entangling region at the holographic boundary \cite{Ryu:2006ef,Ryu:2006bv},
\begin{equation}\label{eq:SEEdef}
S_{EE} = \frac{\mathbb{A}_\mathrm{min}}{4 G_N}
\end{equation} 
where $G_N$ is the five dimensional Newton's constant. This prescription was subsequently adapted to holographic conformal defects of arbitrary dimension and codimension in \cite{Jensen:2013lxa}, and extended in \cite{Estes:2014hka}. We now briefly summarize some important results from these works, in our present conventions.

Generically, the hypersurface whose area we wish to evaluate wraps the $S^1$ parametrised by $\varphi$ in (\ref{eq:AdS4g}), and resides on a constant time slice. It can thus be parametrised by a function $x(r_{||},r)$. Interestingly, for a spherical entangling region of radius $R$ centred on the interface, a proof is provided in \cite{Jensen:2013lxa} that the minimal surface corresponds to the solution
\begin{equation}
x(r_{||},r)^2 = R^2 - r_{||}^2,
\end{equation}
which is to say a surface independent of $r$. The area of this minimal surface is then\footnote{In what follows, unless otherwise stated, we will take the AdS$_4$ radius $\ell = 1$ for convenience.}  given by the integral expression
\begin{equation}\label{eq:Amin}
\mathbb{A}_{\mathrm{min}}= 2\pi R\int_{\mathfrak{x}_c}^R\frac{\dd x}{x^2}\int_{\mathfrak{r}_-}^{\mathfrak{r}_+}\dd r \,e^{2A}.
\end{equation}

While simple in appearance, the expression (\ref{eq:Amin}) requires some further explanation to employ. In particular, there is some subtlety introduced in the limits of integration $\mathfrak{x}_c$ and $\mathfrak{r}_\pm$. These limits introduce a cut-off scale $\epsilon$ which regulates the divergences in the area integral that arise from contributions near the asymptotic boundary.  

A convenient prescription for implementing this cut-off is detailed in \cite{Jensen:2013lxa, Estes:2014hka}. The fundamental idea is to first bring the near boundary metric to a canonical Fefferman-Graham (FG) form where possible. In such a region, the FG radial coordinate (which we henceforth call $\xi$) holographically encodes the energy scale of the dual field theory in the familiar way. Thus, implementing a cut-off along the surface $\xi = \epsilon$ geometrically encodes the regularization of UV divergences in the entanglement entropy coming from modes very near the boundary of the entangling region.

In the FG coordinates we employ, an asymptotically AdS$_5$ region of the spacetime with radius $L$  takes the form
\begin{equation}
\dd s^2 = \frac{L^2}{\xi^2}\left[g_{||}\left(\frac{\xi}{y_3} \right)\left(\dd t^2 - \dd r_{||}^2 - r_{||}^2\dd\varphi^2 \right) -g_\perp\left(\frac{\xi}{y_3} \right)\dd y_3^2\right]-L^2 \frac{\dd \xi^2}{\xi^2},
\end{equation}
where we further impose that in the limit $\xi\to 0$ the functions $g_{||}\to1$ and $g_\perp \to 1$.
A straightforward exercise reveals that the change in coordinates necessary to bring an asymptotically AdS$_5$ metric in the coordinates of \ref{eq:anstz} to this form is given in general by (inverting)
\begin{equation}
\xi = x \,\mathfrak{G}^\pm(r) \qquad \mathrm{and}\qquad y_3 = x\, \mathfrak{F}^\pm(r)
\end{equation}
where 
\begin{equation}\label{eq:ccFG}
\mathfrak{G}^\pm(r) = \exp\left(\mp\int\dd r' e^{-A}\sqrt{\frac{e^{2A}}{L_\pm^2}-1}\,\right),\qquad   \mathfrak{F}^\pm(r) = \exp\left(\mp\int \dd r' \frac{e^{-A}}{\sqrt{\frac{e^{2A}}{L_\pm^2}-1}} \right).
\end{equation}
The $\pm$ has been used to denote an asymptotically AdS$_5$ boundary obtained as $r\to\pm\infty$.

Evaluating (\ref{eq:ccFG}) at $\xi = \epsilon$ and employing the asymptotic form of the metric function $A(r)$ (see e.g. appendix \ref{app:FG}) allows one to obtain expressions for $\mathfrak{r}_\pm(\epsilon/x)$ perturbatively near the FG boundary. In particular, we find
\begin{equation}\label{eq:rcut}
\mathfrak{r}_\pm\left(\frac{\epsilon}{x} \right) = \pm L_\pm \left[ \ln\left(\frac{2x}{\epsilon} \right)-A_0^\pm+\ln\left(\frac{L_\pm}{2}\right) -\frac{1}{4}\left(\frac{\epsilon}{x} \right)^2+\ldots\right]
\end{equation}
where the omitted terms, higher order in $\epsilon/x$, will play no role in what follows. Finally, we will  cut off the $x$-integral at $\mathfrak{x}_c = \epsilon$, in accordance with the prescription advocated for in \cite{Estes:2014hka}. 

\section{RG interface entanglement entropy}\label{sec:4d}

We now bring these results to bear on the holographic solutions of interest. As a preliminary exercise we will first evaluate the vacuum entanglement entropy for a spherical entangling region of radius $R$ in the vacuum of a four dimensional SCFT in the holographic limit. 

Such vacua are holographically dual to AdS$_5$ supergravity solutions. In our conventions, these are constant scalar solution with a metric warp factor of the form
\begin{equation}
e^{A} = L \cosh \frac{r}{L},
\end{equation}
which behaves near the boundary at $r\to\pm\infty$ as
\begin{equation}
A = \pm \frac{r}{L} +  \ln\frac{L}{2} + \ldots
\end{equation}
implying that the cut-offs $\mathfrak{r}_\pm$ in such a background evaluate to
\begin{equation}\label{eq:vCut}
\mathfrak{r}_\pm\left(\frac{\epsilon}{x} \right) = \pm L\, \mathrm{arccosh}\left(\frac{x}{\epsilon} \right)=\pm L \left[ \ln\left(\frac{2x}{\epsilon} \right)-\frac{1}{4}\left(\frac{\epsilon}{x} \right)^2+\ldots\right].
\end{equation}

For the vacuum solution, in which an FG radial coordinate can be defined asymptotically close to the entirety of the boundary, continuity of the cut-off surface implies that the exact result in the first equality of (\ref{eq:vCut}) must be used in the limits of integration for the radial integral in (\ref{eq:Amin}). Evaluating this integral then gives the sphere entanglement entropy in the holographic CFT vacuum as
\begin{equation}\label{eq:vacSEE}
S_{EE}^{\mathrm{CFT}} = \frac{\pi}{2}\frac{L^3}{G_N}\left[\frac{R^2}{\epsilon^2}-\ln\left(\frac{2 R}{\epsilon} \right)-\frac{1}{2} \right],
\end{equation}
where terms vanishing in the $\epsilon\to 0$ limit have been dropped.

The dimensionless combination $L^3/G_N$ can be written entirely in terms of field theory quantities. For example, for the AdS$_5$ solution dual to the $\mathcal{N}=4$ SYM vacuum, 
\begin{equation}\label{eq:L3oG}
\frac{L^3}{G_N} = \frac{2}{\pi}N^2,
\end{equation}
with $N$ the rank of the theory's gauge group in the large $N$ limit.

As we shall see, although the LS side of the interface is undeformed by any couplings to relevant operators, the same is not generically true on the $\mathcal{N}=4$ SYM side. Accordingly, the ambient theory's contribution to the sphere entanglement entropy  will contain a contribution of the form (\ref{eq:vacSEE}) from the LS side. However, a more careful analysis will be needed to identify and subtract the contribution from the ambient theory's $\mathcal{N}=4$ SYM side. We will return to this task shortly.

Moreover, since the family of holographic RG interface solutions are only known numerically, special care must be taken to ensure that $\mathcal{C}^I$ is properly extracted from the full sphere entanglement entropy. In particular, unaccounted for numerical error could prevent a precise subtraction of the ambient theory contributions, which in turn might yield a result which in fact diverges as the UV cut-off is removed. We thus find it extremely beneficial to employ a semi-analytical approach in which these divergences can be handled directly. 

\subsection{Computational Approach}\label{sec:meth}
Our method is straightforward. We work in units such that the maximally symmetric AdS$_5$ radius $L = 1$, and thus (in accordance with the discussion in section \ref{sec:sols}) we have $L_+ = 1$ and $L_- =3/2^{5/3}$. The linear BPS mode (\ref{eq:BPSm}) used to construct the holographic interface solutions does not modify the warp factor at linear order, so all solutions in our family are characterized asymptotically by $A_0^- = \ln 3/2^{8/3}$. From (\ref{eq:rcut}) we then obtain
\begin{equation}
\mathfrak{r}_- = -\frac{3}{2^{5/3}}\left[\ln\left(\frac{2 x}{\epsilon}\right)-\frac{1}{4}\left(\frac{\epsilon}{x} \right)^2\right], \qquad  \mathfrak{r}_+ =\left[\ln\left(\frac{2 x}{\epsilon}\right)-A_0^+-\ln 2-\frac{1}{4}\left(\frac{\epsilon}{x} \right)^2 \right].
\end{equation}
The constant $A_0^+$ must be read off from the asymptotic behaviour of the warp factor solution by solution. We find it convenient to do so by performing a linear regression at large values of $r$. 

In order to circumvent some of the technical difficulties inherent to the numerical ambient theory subtraction, we use a combination of analytic approximation and numerical integration to evaluate (\ref{eq:Amin}). More specifically, we first write this integral expression as
\begin{equation}\label{eq:AminN}
\mathbb{A}_{\mathrm{min}} = 2\pi R\int_\epsilon^R\frac{\dd x}{x^2}\left[\mathcal{I}^+ + \int_{r_{0}^-}^{r_{0}^+} \dd r \,e^{2A} -  \mathcal{I}^-\right]
\end{equation}
where
\begin{equation}\label{eq:FGint}
\mathcal{I}^\pm \equiv \int_{{r_0}^\pm}^{\mathfrak{r}_\pm}\dd r\,e^{2A}.
\end{equation}
Note that, importantly, all of the $x$ dependence of the $r$-integral has been shuffled into the $\mathcal{I}^\pm$. At this juncture, the integration limits $r_0^\pm$ are completely arbitrary---their dependence in each of the three terms in the square brackets of (\ref{eq:AminN}) is removed upon taking the (signed) sum.

\begin{figure}[h]
\centering
\includegraphics[scale=0.6]{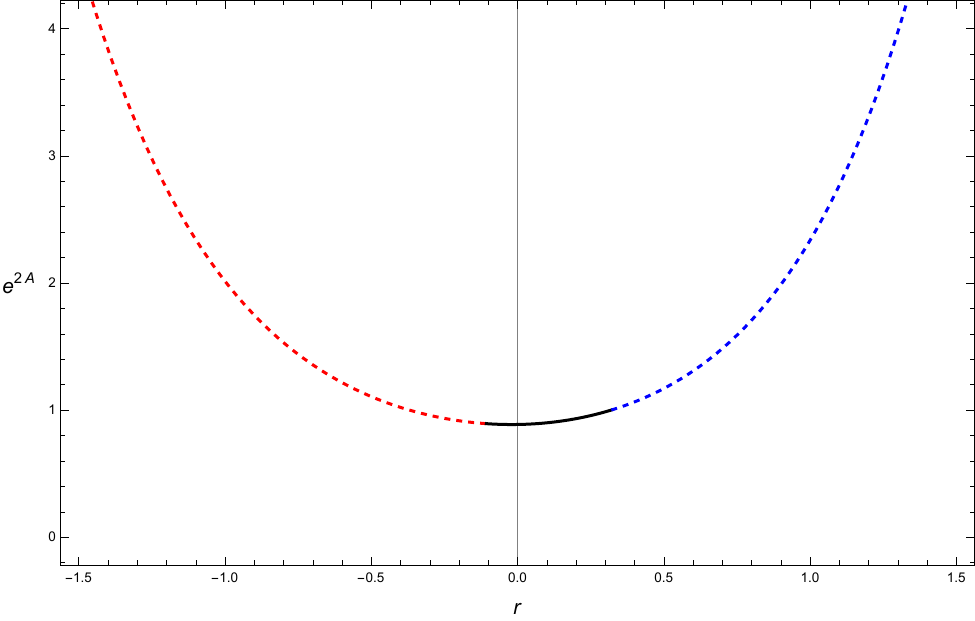}
\caption{The dashed colored curves indicate the domain of $r$ such that an FG asymptotic expansion of the solution can be performed. The red dashed curve illustrates the feasibility of such an expansion around the LS boundary, while the blue dashed curve denotes the FG regime corresponding to the $\mathcal{N}=4$ boundary. Along the black segment, inequality (\ref{eq:FGcon}) is not satisfied, and accordingly this region is not accessible from FG expansion.} \label{fig:FGRGI}
\end{figure}

We next approximate the integrand of the $\mathcal{I}^\pm$ integrals, replacing the integrand by the analytic form obtained via FG expansion near the corresponding asymptotic boundary. In order for this approximation to make sense, equation (\ref{eq:ccFG}) requires that the $r_0^\pm$ must be chosen such that the following inequality is satisfied:
\begin{equation}\label{eq:FGcon}
e^{2A(r_0^\pm)} > L_\pm^2.
\end{equation}
For a representative RG interface solution, that with $\phi_{(s)}=0$,  the domain of $r$ such that inequality (\ref{eq:FGcon}) is satisfied is shown in figure \ref{fig:FGRGI}.

In practice, $r_0^\pm$ is chosen such that the numerically obtained background solution evaluated at $r_0^\pm$ yields an integrand which agrees with that of the analytic FG approximation of the integrand to order $\mathcal{O}(e^{-r/L_\pm})^0$ to an accuracy of one part in a thousand.\footnote{Note that continuity of the integrand together with its first derivative at $r_0^\pm$ is essentially the Israel junction condition for joining the numerically generated spacetime to its near boundary analytic behavior along the (tensionless) $r=r_0^\pm$ hypersurface.}. We verify the $r_0^\pm$ independence of our results by re-running our numerical algorithm over many values of $r_0^\pm$ subject to these accuracy requirements and inequality (\ref{eq:FGcon}).

The advantage of this approach is considerable. It allows one analytic control of the near boundary divergences in the minimal area integral, and thus permits a precise means of ambient theory subtraction. It is helpful to see this in action. To do so, we assume that in the FG regime, the warp factor $A$ has an asymptotic expansion of the form (see appendix \ref{app:FG} for further details)
\begin{equation}
A = \pm\frac{r}{L_\pm}+ A_0^\pm + A_2^\pm\, e^{\mp 2r/L_\pm} + \ldots
\end{equation}
where the subleading terms contained in the ellipses will play no role in the immediate discussion.

Using this asymptotic form, it is straightforward to evaluate the integrals in (\ref{eq:FGint}). One obtains
\begin{multline}\label{eq:Ipm}
\mathcal{I}^\pm = \pm\frac{L_\pm^3}{2}\left[\left(\frac{x}{\epsilon} \right)^2 - \frac{1}{2} \right]\\\pm
2 L_\pm A_2^\pm e^{2A_0^\pm}\left[\ln\left(\frac{2x}{\epsilon} \right)-A_0^\pm+\ln\left(\frac{L_\pm}{2} \right)-\frac{1}{4}\left(\frac{\epsilon}{x}\right)^2 \right] - \mathcal{J}^\pm\left[r_0^\pm\right]+\ldots
\end{multline}
where 
\begin{equation}
\mathcal{J}^\pm\left[r_0^\pm \right] \equiv e^{2A_0^\pm}\left(2A_2^\pm r_0^\pm \pm \frac{L_\pm}{2}
e^{\pm2 r_0^\pm/L_\pm} \right).
\end{equation}
Observe that all the $r_0^\pm$ dependence in the $\mathcal{I^\pm}$ integral is contained in the $\mathcal{J}^\pm$. As explained above, this dependence must drop out when the sum in (\ref{eq:AminN}) is performed. Moreover, from (\ref{eq:Ipm}) one obtains analytically the leading $x$-dependent contributions to the integral in the limit that $\epsilon\to 0$.

It is now trivial to perform the remaining $x$-integral to obtain the area of the minimal surface. The result is
\begin{multline}\label{eq:AminFG}
\mathbb{A}_{\mathrm{min}} = \pi\left(L_+^3 + L_-^3 \right)\left(\frac{R}{\epsilon} \right)^2-4\pi\left(L_+ A_2^+ e^{2A_0^+}+L_- A_2^- e^{2A_0^-} \right)\ln\left(\frac{2R}{\epsilon} \right) + s_1 \left(\frac{R}{\epsilon}\right)\\
+ \frac{\pi}{2}\left(L_+^3 + L_-^3 \right) - 4\pi \left[L_+ A_2^+ e^{2A_0^+}\left(1+\ln\left(\frac{L_+}{2}\right) -A_0^+\right) +L_- A_2^- e^{2A_0^-}\left(1+\ln\left(\frac{L_-}{2}\right) -A_0^-\right)\right]\\
+2\pi\left(\mathcal{J}^+\left[r_0^+ \right]-\mathcal{J}^-\left[ r_0^-\right] \right)-2\pi \int_{r_0^-}^{r_0^+} \dd r \,e^{2A}.
\end{multline}
The terms divergent in the $\epsilon\to 0$ limit are written on the first line of (\ref{eq:AminFG}). These include a linearly divergent term with coefficient $s_1$. Although $s_1$ can also be written in terms of FG expansion coefficients, as it is a somewhat complicated and non-universal (cut-off dependent) quantity we refrain from doing so here. Explicit expressions can be found in \cite{Estes:2014hka}.

Given (\ref{eq:AminFG}), it is clear that in order to define a universal quantity, a minimal subtraction scheme must at the very least remove the first line of (divergent) terms. The second line, however, contains terms finite in the $\epsilon\to 0$ limit which are also characteristic of the ambient theory. To see this, consider for example the contribution coming from the LS side of the interface. Using the results of appendix \ref{app:FG}, in particular
\begin{equation}
A_0^- = \ln\left(\frac{L_-}{2}\right) \qquad \mathrm{and} \qquad A_2^- = 1,
\end{equation}
one finds that the contribution from the ``$-$'' side to the first two lines yields exactly (one half) the LS vacuum result (\ref{eq:vacSEE}), modulo the linear divergence which only appears due to the presence of the interface. Similar comments apply also to the ``+'' side of the interface. Notably however, the divergences on this side are {\it not} those of the $\mathcal{N}=4$ SYM vacuum, but rather those of the $\mathcal{N}=4$ SYM theory deformed by the coupling to the supersymmetric mass terms. As demonstrated in appendix \ref{app:FG}, these terms explicitly depend on $\phi_{(s)}$. In particular, we note that $A_2^+$ encodes the leading departure from the $\mathcal{N}=4$ SYM vacuum
\begin{equation}
A_2^+ = \frac{1}{4}\left( e^{-2A_0^+}-\frac{2}{3}\phi_{(s)}^2\right),
\end{equation}
and that to this order the backreaction from the scalar sector is pure ``source''---the coefficients governing the subleading fall-offs do not appear. It is thus natural to associate the `+' terms appearing explicitly in the first and second lines of (\ref{eq:AminFG}) with the ambient theory's contribution to the sphere entanglement entropy.

In so far as we are interested in defining a universal quantity characteristic of the defect, we will thus employ a prescription in which the finite contribution on the second line of (\ref{eq:AminFG}) is also subtracted from the sphere entanglement entropy. In other words, we will define the universal part of the defect contribution to the sphere entanglement entropy as
\begin{equation}\label{eq:NCI}
\mathcal{C}^I \equiv \frac{\pi}{2G_N} \mathcal{J}^+\left[r_0^+\right] - \frac{\pi}{2G_N} \int_{r_{0}^-}^{r_{0}^+} \dd r \,e^{2A} -\frac{\pi}{2 G_N}\mathcal{J}^-\left[r_0^-\right].
\end{equation}
For the RG interfaces studied in this work, we are thus faced with the task of evaluating (\ref{eq:NCI}) with 
\begin{align}
\mathcal{J}^+\left[r_0^+ \right] & = e^{2A_0^+}\left[\frac{1}{2}\left(1-\frac{2}{3}\phi_{(s)}^2 \right) r_0^+e^{-2A_0^+} + \frac{1}{2}
e^{2 r_0^+} \right], \\
\mathcal{J}^-\left[r_0^- \right] & = \frac{L_-^2}{4}\left[2 r_0^- - \frac{L_-}{2}
e^{-2 r_0^-/L_-} \right] \qquad \mathrm{where} \qquad L_- = 3/2^{5/3}
\end{align}
on our numerically constructed supersymmetric solutions. Other prescriptions for defining the interface contribution to the entanglement entropy are explored in appendix \ref{app:schemes}.

\subsection{Results}\label{sec:res}
\begin{figure}[h]
\centering
\includegraphics[scale=1.4]{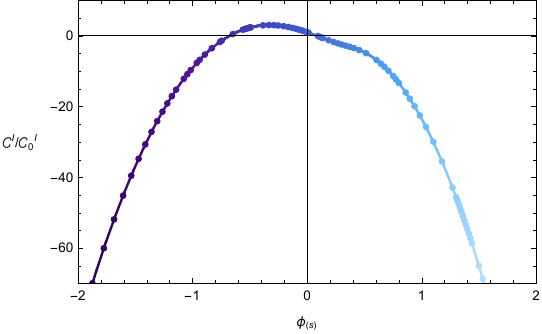}
\caption{The interface contribution to the sphere entanglement entropy for the RG interfaces, $\mathcal{C}^I$, normalized by its value for the source free RG interface $\mathcal{C}_0^I$. The color coding is correlated with that of figure \ref{fig:ImzvsRez}.}\label{fig:numCI}
\end{figure}

Our results are summarized in figure \ref{fig:numCI}. There, the interface contribution to the sphere entanglement entropy, $\mathcal{C}^I$, has been normalized by its value at the source free RG interface:
\begin{equation}\label{eq:CI0}
\mathcal{C}^I_0 \approx 0.040\cdot \frac{2}{\pi}N^2.
\end{equation}
In writing (\ref{eq:CI0}) we have used the entry (\ref{eq:L3oG}) from the holographic dictionary to write our result in terms of field theory quantities. 

\begin{figure}[h]
\centering
\includegraphics[scale=0.85]{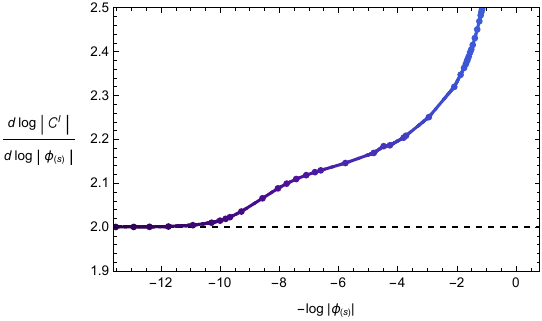}
\includegraphics[scale=0.85]{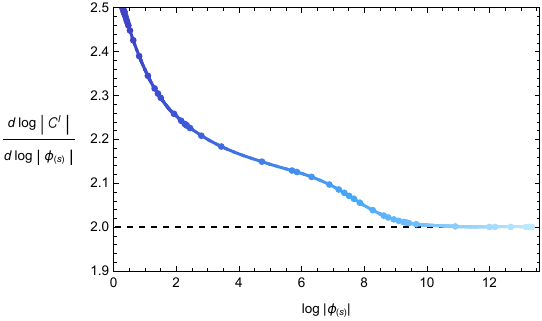}
\caption{The logarithmic derivative of $\mathcal{C}^I$ with respect to the source $\phi_{(s)}$. The color coding is correlated with that of figure \ref{fig:ImzvsRez}. The quadratic behavior at large $|\phi_{(s)}|$ is clearly visible as $\phi_{(s)}\to \pm\infty$.}\label{fig:dCdphi}. 
\end{figure}

Evidently, $\mathcal{C}^I$ changes sign across this family of RG interfaces, vanishing twice. Similar behavior has been previously observed in other defect field theories, e.g. \cite{Jokela:2025qac,Conti:2025wwf,Kobayashi:2018lil}. When $\phi_{(s)}$ becomes large in magnitude, we find that $\mathcal{C}^I$ diverges towards negative infinity. An analysis of this asymptotic behavior suggests that this divergence is power law, 
\begin{equation}
\mathcal{C}^I\left(|\phi_{(s)}|\to\infty\right) \sim  |\phi_{(s)}|^\gamma \qquad \mathrm{with} \qquad \gamma \approx 2.
\end{equation}
This scaling behavior is readily observed in the asymptotics of the logarithmic derivative of $\mathcal{C}^I$, shown in figure \ref{fig:dCdphi}.

The large source limit is of particular interest because, as shown in \cite{Arav:2020asu}, when the magnitude of $\phi_{(s)}$ diverges the RG interfaces approach one of two limiting solutions. As $\phi_{(s)}$ diverges towards {\it negative} infinity, the bulk solution increasingly resembles a ``junction'' type geometry in which the AdS$_5^{\mathrm{LS}}$ vacuum is glued to the bulk solution dual to the Poincar\'e invariant RG flow between $\mathcal{N}=4$ SYM and the LS$^+$ SCFT along a constant $r$ hypersurface. As $\phi_{(s)}$ diverges towards {\it positive} infinity, the RG interface solution instead resembles a junction solution in which the Poincar\'e invariant RG flow between $\mathcal{N}=4$ SYM and the LS$^-$ SCFT is glued to a Janus solution interpolating between the LS$^+$ and LS$^-$ boundaries.

It would be interesting to pursue this perspective on the large $\phi_{(s)}$ limit further. In particular, it is plausible that it could be leveraged to obtain analytic control over the interface contribution to the entanglement entropy in these asymptotic regimes. While we will not follow this direction in the present work, we note that similar ``fragmentation'' limits of holographic interface solutions have also been studied in \cite{Ghodsi:2022umc}. In appendix \ref{app:RGscaling}, we explore this large deformation regime in further detail, providing suggestive numerical evidence that the observed power law behavior of $\mathcal{C}^I$ is correlated with the appearance of a growing region of the bulk geometry which tracks the known {\it Poincar\'e invariant} holographic RG flow between $\mathcal{N}=4$ SYM and the LS fixed point theory.

\section{On-shell action and stress tensor expectation values for RG interfaces}\label{sec:obs}
It is natural to wonder whether or not the defect quantity $\mathcal{C}^I$ appears in other defect field theory observables. An immediate candidate, motivated in part by the results of \cite{Casini:2023kyj}, is the defect contribution to the on-shell action, $S_D$. The holographic computation of the renormalized on-shell action for the RG interface solutions is a largely mechanical exercise aided considerably by the results collected in appendix B of \cite{Arav:2020obl}. Here we highlight some important features of this analysis, relegating additional details to appendix \ref{app:OS}.

Our starting point is the observation\footnote{Similar considerations apply also to holographic conformal defects of other dimension and co-dimension. See for example footnote 32 of \cite{Arav:2024exg}, in concert with the entropic central charge formula employed in \cite{Conti:2025wwf,Conti:2025wyj}.}  that the bulk contribution to the on-shell action can be written
\begin{equation}
S^{\mathrm{OS}}_{\mathrm{bulk}} = -\frac{1}{8\pi G_N}\int \dd^4 x \sqrt{|\gamma|} A'\Big |^{\mathfrak{r}_+}_{\mathfrak{r}_-} - \frac{3}{8\pi G_N}\cdot\mathrm{vol}\left(\mathrm{AdS}_4 \right)\int^{\mathfrak{r}_+}_{\mathfrak{r}_-} \dd r\,e^{2A}.
\end{equation}
In writing this expression, we have implicitly assumed that we are working in a conformal frame in which the boundary is AdS$_4$. In particular, unlike in previous sections, here we will regulate the action along cut-off hypersurfaces at constant $r = \mathfrak{r}_\pm$. While the first term is a boundary term,  the second involves an integral over the bulk. In fact, this bulk integral is the same as the radial integral appearing in (\ref{eq:Amin}), and it is thus perhaps not surprising that the on-shell action can be written in terms of the defect contribution to the sphere entanglement entropy. 

After combining the bulk on-shell contribution with the divergent and finite counterterms, the holographically renormalized on-shell action for any RG interface in the family studied here evaluates to
\begin{multline}\label{eq:SosP}
S^{\mathrm{OS}} = \frac{\mathrm{vol}\left(\mathrm{AdS}_4 \right)}{4\pi G_N}\Bigg[\frac{3 G_N}{\pi}\mathcal{C}^I +\phi_{(s)}\alpha_{(v)}+\frac{3}{16}\left(1+4\delta_{R^2} -28\delta_{\Delta R^2}\right)
-\left(\delta_{\alpha}+\delta_{R\phi^2(1)} \right)\phi_{(s)}^2 \\+ \frac{3}{16}L_-^3\left(1+4\tilde{\delta}_{R^2} -28\tilde{\delta}_{\Delta R^2} \right)\Bigg]
\end{multline}
with $L_- = 3/2^{5/3}$. In four dimensions, this finite contribution to the on-shell action is scheme dependent.  This fact is reflected in the appearance of the $\delta$ and $\tilde{\delta}$, which are coefficients of finite counterterms in the holographic renormalization scheme. While it is conceivable that (super)symmetry considerations impose further constraints relating them, we are currently unaware of any additional relationships between these couplings.

We now demonstrate that this on-shell action can be further expressed in terms of data governing the one-point function of the dCFT's stress tensor. This one point function can be computed in the standard way, varying the on-shell action with respect to the boundary metric. The isometries of the solution ensure that the expectation value takes the form
\begin{equation}
\langle \mathcal{T^\pm}_{ab} \rangle\dd x^a \dd x^b = a_T^\pm \,\dd s^2\left(\mathrm{AdS}_4 \right)
\end{equation}
on either side of the interface. Here $a_T^\pm$ is a constant that generically can depend on conformal couplings deforming the ambient CFTs. As an illustrative example, consider the `$+$' side of the interface, corresponding to the deformed $\mathcal{N}=4$ theory. The trace Ward identity gives 
\begin{equation}\label{eq:TN4P}
a_T^{\mathcal{N}=4} = \frac{1}{4}\Big(\mathcal{A}^{\mathcal{N}=4}-\phi_{(s)}\langle\mathcal{O}_\phi \rangle-2\alpha_{(s)}\langle\mathcal{O}_\alpha \rangle -2\beta_{(s)}\langle\mathcal{O}_\beta \rangle \Big),
\end{equation}
where $\mathcal{A}^{\mathcal{N}=4}$ is the conformal anomaly for the deformed $\mathcal{N}=4$ theory on AdS$_4$. Explicit expressions for both the anomaly and the scalar expectation values can be found in \cite{Arav:2020obl}. Evaluating them in the current setting yields
\begin{equation}\label{eq:aTNis4P}
a_T^{\mathcal{N}=4} = \frac{1}{4\pi G_N}\left[\frac{3}{16}-\phi_{(s)}\alpha_{(v)}+\frac{1}{4}\left(1+4\delta_\alpha+2\delta_{R\phi^2(1)} \right)\phi_{(s)}^2 \right]
\end{equation}
and
\begin{equation}\label{eq:aTLSP}
a_T^{\mathrm{LS}} = \frac{L_-^3}{4\pi G_N}\left(\frac{3}{16} \right).
\end{equation}
On the LS side of the interface, which is undeformed by any couplings for scalar operators, the non-zero $a_T$ is entirely a consequence of the Casimir energy for the theory on AdS$_4$. We note that under Weyl transformation to flat space, $\langle \mathcal{T}^{\mathrm{LS}}_{ab} \rangle$ vanishes by symmetry constraints, but $\langle \mathcal{T}^{\mathcal{N}=4}_{ab} \rangle$ generically does not due to the presence of non-zero $\phi_{(s)}$. 

Specifically, for a codimension one defect in flat space, conformal invariance together with conservation and tracelessness of the stress tensor do not allow a non-vanishing stress tensor one-point function away from the interface \cite{Billo:2016cpy}. These arguments do not directly apply on the  deformed $\mathcal{N}=4$ SYM side of the interface. There, the stress tensor expectation value can be nonzero, consistent with the presence of source dependent terms in the stress tensor Ward identities \cite{Arav:2020obl}.

Comparing (\ref{eq:TN4P}) and (\ref{eq:aTLSP}) to (\ref{eq:SosP}), we observe that if one further implements a scheme in which the finite counterterms obey
\begin{align}
\delta_{R^2} & = -\frac{1}{2}+7\delta_{\Delta R^2} \nonumber\\
\delta_{R\phi^2(1)} & = \frac{1}{2}\label{eq:fcts} \nonumber\\
\tilde{\delta}_{R^2} & = -\frac{1}{2}+7\tilde{\delta}_{\Delta R^2}
\end{align}
the on-shell action then takes the suggestive form
\begin{equation}
S^{\mathrm{OS}} = \mathrm{vol}\left(\mathrm{AdS}_4 \right)\Bigg[\frac{3}{4\pi^2}\mathcal{C}^I -\left(a_T^{\mathcal{N}=4}+a_T^{\mathrm{LS}}\right) \Bigg].
\end{equation}
Independent of the values taken by the finite counterterm coefficients, it is interesting to wonder whether or not it is possible to define scheme independent ``subtracted'' quantities $I_D$ and $a_D$ which capture some universal physics inherent to the defect contributions to the on-shell action and stress tensor one-point function. Such a disentangling is non-trivial, in the sense that the existence of the holographic RG interfaces in the family studied here depends crucially on the details of the mass deformations of the $\mathcal{N}=4$ theory controlled by $\phi_{(s)}$. 

Nonetheless, we are tempted to adopt a subtraction prescription analogous to that employed in the definition (\ref{eq:NCI}). Specifically, we subtract from (\ref{eq:SosP}) and the sum of (\ref{eq:aTNis4P})and (\ref{eq:aTLSP}) the terms characteristic of the undeformed ($\phi_{(s)}=0$) vacuum, as well as the terms quadratic in $\phi_{(s)}$. Such terms would be present for any two putative states in the theory deformed by the same supersymmetric mass term, and thus could not be used to distinguish between them. 

Applying this prescription\footnote{Alternatively, we could arrive at the same definitions by employing (\ref{eq:fcts}) and subsequently fixing the remaining finite counterterm to $\delta_\alpha = -1/2$. We will refrain from imposing this relationship, as we are not currently aware of any compelling argument that would privilege this scheme.} and analytically continuing to Euclidean space such that the on-shell action becomes $S^\mathrm{OS}\to-I^{OS}$, we define
\begin{equation}
S_D = -I_D \equiv  \frac{\mathrm{vol}\left(\mathbb{H}^4 \right)}{4\pi G_N}\Bigg[\frac{3 G_N}{\pi}\mathcal{C}^I +\phi_{(s)}\alpha_{(v)}\Bigg] \qquad \mathrm{and} \qquad a_D \equiv -\frac{1}{4\pi G_N}\phi_{(s)}\alpha_{(v)}.
\end{equation}
We further obtain a renormalized volume on  $\mathbb{H}^4$ via the standard prescription (employed in e.g. \cite{Casini:2011kv}). In particular, we foliate $\mathbb{H}^4$ with unit $S^3$ leaves,
\begin{equation}
\dd s^2(\mathbb{H}^4) = \dd \eta^2 + \sinh^2\eta\, \dd \Omega^2_3
\end{equation}
and then compute
\begin{equation}
\mathrm{vol}(\mathbb{H}^4) = \mathrm{vol}\left(S^3 \right)\left[\int_0^{\eta_c} \dd \eta \sinh^3\eta\right]_{\mathrm{finite}} = -\frac{4\pi^2}{3}.
\end{equation}
The instruction for evaluating the integral in square brackets above is to keep only the terms finite in the $\eta_c\to\infty$ limit. 

All together, we find
\begin{equation}
\mathcal{C}^I = I_D +\frac{4\pi^2}{3} a_D.
\end{equation}
This form for the on-shell action is reminiscent of the results of \cite{Casini:2023kyj,Conti:2025wwf,Kobayashi:2018lil,Jensen:2018rxu,Capuozzo:2023fll}, in which the defect contribution to the sphere entanglement entropy is expressed in terms of a defect free energy and an energy term. We note that in the absence of the supersymmetric mass deformation, $a_D$ vanishes and the defect contribution to the sphere entanglement is given entirely by the defect contribution to the on-shell action.

\section{Discussion}\label{sec:dis}
We have computed the entanglement entropy of a spherical region centred on the codimension one interface for a family of RG interface SCFTs in their holographic limit. Unlike the most familiar interface theories, in which the ambient theories on either side of the interface are SCFT vacua (perhaps deformed by spatially homogeneous marginal couplings), here the situation is less symmetric. In particular, on the $\mathcal{N}=4$ SYM side of the interface, a generic member of our family is characterized by a non-vanishing supersymmetric mass term which varies along the direction normal to the interface.

An important output of our calculation is summarized by figure \ref{fig:numCI}. There, we illustrate the dependence of the interface contribution to the entanglement entropy on the deformation parameter $\phi_{(s)}$. In doing so, we have adopted a particular scheme for regularizing the short distance divergences characteristic of entanglement between modes very near the entangling surface. 

Broadly, for such a scheme to yield physically interesting results it must ensure that any computed quantities are insensitive to the details of the short distance cut-off. Here, we ensure that this is the case by performing a background subtraction. In other words, by subtracting divergences characteristic of the ambient theories on either side of the interface, together with an easily identifiable scheme dependent divergence characteristic of the interface itself. This subtraction results in a cut-off independent quantity, but is not unique.

Nonetheless, we expect that our calculations reveal universal properties of entanglement in these RG interfaces. In particular, by repeating our computations in various alternative schemes (some of which are discussed in appendix \ref{app:schemes}) we find that the cut-off independent part of the interface entanglement entropy can in general vanish, change sign, and attain local extrema along this family of solutions. 

Moreover, in the limit that the deformation parameter becomes very large in magnitude, the interface contribution to the entanglement entropy $\mathcal{C}^I$ attains a power law dependence on $\phi_{(s)}$, diverging quadratically. This singular limit is of particular interest, as the dual bulk solutions asymptote in this case to ``junction'' type geometries in which the solution dual to the Poincar\'e invariant RG flow between $\mathcal{N}=4$ SYM and the LS SCFT is glued to either an AdS$_5^{\mathrm{LS}}$ or LS$^+$/LS$^-$ Janus region. It would be interesting to understand this asymptotic regime in more detail, specifically in the context of the interface entanglement entropy.

By carefully applying the machinery of holographic renormalization to these RG interface solutions, we demonstrate an interesting relationship between the renormalized on-shell action of the interface SCFT and the cut-off independent part of the sphere entanglement entropy in these systems. Holographically, this follows from an application of the Wald formalism to the computation of the on-shell bulk action in the conformal interface ansatz. This perspective can be used to derive analogous relationships between the defect contribution to the on-shell action and its contribution to the sphere entanglement entropy for holographic conformal defects of other codimension.

The holographic renormalization introduces an ambiguity in the computation of field theory observables through the presence of scheme dependent coefficients for finite boundary counterterms. While supersymmetry can be used to constrain some of these coefficients, it is not presently known to what extent it constrains the rest. We find a suggestive relationship between some of the un-fixed coefficients that allows us to write the on-shell action in terms of both the interface contribution to the entanglement entropy as well as an ``energy term'', when the ambient theories are placed on AdS$_4$.

An obvious open question is whether or not the quantity $\mathcal{C}^I$ we have computed for these RG interfaces obeys any sort of monotonicity along RG flow trajectories. In the context of dCFTs, it is natural to distinguish between two types of RG flow. In the first, in which the deformation driving the RG flow is localized to the interface, the results of \cite{Casini:2023kyj} plausibly imply that $\mathcal{C}^I$ is an RG monotone for the RG interface in which $\phi_{(s)}$ vanishes (and thus $\mathcal{C}^I = I_D$). It is interesting to wonder whether $I_D$ continues to play this role when the interfaces are characterized by non-vanishing $\phi_{(s)}$. 

For the second type of defect RG flow, in which the deformation is applied to the ambient theory (i.e. away from the interface), far less is known. In \cite{Conti:2025wwf}, results for holographic conformal defects of other codimension suggest that the analogous quantity is non-monotonic under bulk RG flow. Similar results have been reported in non-holographic theories as well, in \cite{Herzog:2019rke,Shachar:2024ubf}. 

Our family of RG interfaces leaves little opportunity to address such questions directly. In particular, each member of the family is characterized by a different value of the deformation parameter, and thus each provides the dual of a distinct dCFT fixed point. The solution in which this parameter, $\phi_{(s)}$, vanishes is a privileged member of the family, in the sense that it provides the dual of an interface between undeformed $\mathcal{N}=4$ SYM and LS vacua. 

In this context, one might wonder whether or not this dCFT flows to the LS$^+$/LS$^-$ Janus solution if the $\mathcal{N}=4$ theory is deformed on the half space by the spatially homogeneous mass term  which drives the Poincar\'e invariant flow between the $\mathcal{N}=4$ and LS theories. If so, the existence of such a flow might in principle be used to understand holographically the behavior of $\mathcal{C}^I$ at the UV and IR of a bulk RG interface RG flow. As the  LS$^+$/LS$^-$ Janus solution is currently beyond the reach of our holographic computations, we will refrain from further speculation along this direction and leave more careful consideration of the RG flow  behavior of $\mathcal{C}^I$ for RG interfaces to future work. 

\section*{Acknowledgements}
We are indebted to Igal Arav, Matthew Cheung, Jerome Gauntlett, and Matthew Roberts for collaboration on related topics and enlightening conversation. We also thank Ignacio Carre\~no Bolla, Carlos Hoyos, Elias Kiritsis and Ricardo Stuardo for useful discussions. EA would like to thank the Crete Center for Theoretical Physics and King's College London for the kind hospitality while some parts of this work were being completed. The work of EA is supported by the Severo Ochoa fellowship PA-23-BP22-170 and supported in part by the Spanish national grant MCIU-22-PID2021-123021NB-I00 and PID2024-161500NB-I00. The work of CR is partially supported through the framework of H.F.R.I. call ``Basic research Financing (Horizontal support of all Sciences)'' under the National Recovery and
Resilience Plan ``Greece 2.0'' funded by the European Union--NextGenerationEU (H.F.R.I. Project Number: 15384). 

\begin{appendix}

\section{Fefferman-Graham expansions and asymptotic boundary behavior}\label{app:FG}
Here we catalogue some useful results from a Fefferman-Graham analysis near the asymptotic AdS$_5$ boundaries on either side of the RG interface solutions. As in the body of the text, we work in conventions in which $L_+ = \ell = -\kappa = 1$.
\subsection{The $\mathcal{N}=4$ SYM boundary}
We have constructed our solutions such that as $r \to +\infty$ the background asymptotes to the maximally symmetric AdS$_5^{\mathcal{N}=4}$ dual to $\mathcal{N}=4$ SYM. The general form of the near boundary expansion for the bulk fields near this boundary is
\begin{align}\label{eq::FGN4exp}
    A&=r + A_0^+ + A_2^+ e^{-2r}+\left(A_{42}^+ r^2 + A_{41}^+r + A_{40}^+ \right)e^{-4r}  +\cdots\nonumber\\ 
    \phi &= \phi_{(s)}e^{-r}+\phi_{(v)} e^{-3r}+\phi_{31} r e^{-3r} +\cdots \nonumber\\ 
    \alpha &=\left(\alpha_{(s)} r+\alpha_{(v)}\right)e^{-2r}+\left(\alpha_{42}r^2 +\alpha_{41}r+\alpha_{40} \right)e^{-4r} +\cdots \nonumber\\ 
    \beta &= \left(\beta_{(s)} r +\beta_{(v)} \right)e^{-2r}+\left(\beta_{42}r^2+\beta_{41}r+\beta_{40} \right)e^{-4r} +\cdots
\end{align}
The BPS equations imply the source relations
\begin{equation}
    \beta_{(s)}=-\frac{2}{3}\phi_{(s)}^2 \ ,\quad \alpha_{(s)}=\phi_{(s)}
\end{equation}
as well as, for example, 
\begin{align}
 A_2^+& = \frac{e^{-2A_0^+}}{4}-\frac{\phi_{(s)}^2}{6}\nonumber\\
    A_{42}^+&= -\frac{e^{-2A_0^+}\phi_{(s)}^2}{6}-\frac{4\phi_{(s)}^4}{9}\nonumber\\
    A_{41}^+&= \frac{1}{9} \left(\alpha _v \left(-8 \phi _s^3 e^{A_0^+}-3 \phi _s e^{-A_0^+}\right)-\frac{9}{4} \phi _s^2 e^{-2 A_0^+}+2 \phi _s^4\right)\nonumber \\
    A_{40}^+&= \frac{1}{288} \bigg(8 \phi _s \alpha _v e^{-A_0^+} \left(8 \phi _s^2 e^{2 A_0^+}-9\right)-16 \alpha _v^2 \left(8 \phi _s^2 e^{2 A_0^+}+3\right)\nonumber\\ & \qquad  -30 \phi _s^2 e^{-2
   A_0^+}-9 e^{-4 A_0^+}+152 \phi _s^4\bigg) \nonumber\\
    \phi_{(v)}&=\frac{1}{24} \left(4 \alpha _v e^{-A_0^+} \left(8 \phi _s^2 e^{2 A_0^+}-3\right)-3 \phi _s e^{-2 A_0^+}+29 \phi _s^3\right)\nonumber \\
    \phi_{31}&= \frac{4 \phi _s^3}{3}-\frac{1}{2} \phi _s e^{-2 A_0^+}.
\end{align}

\subsection{The LS SCFT boundary}
As $r\rightarrow -\infty$, our solutions asymptote to an AdS$_5^{\mathrm{LS}}$ boundary dual to the LS SCFT.  The background can be expanded like
\begin{align}\label{eq::FGLSexp}
    A &= -\frac{r}{L_-}+A_0^-+A_2^- e^{2r/L_-} + A_4^- e^{4r/L_-}\cdots \nonumber\\
   \mathrm{Im}\, z &=2-\sqrt{3}+\zeta e^{r\delta/L_-}+\cdots\nonumber\\
    \mathrm{Re}\, z &= \alpha_1^- e^{r\delta/L_-}e^{r/L_-}+\cdots\nonumber\\
    \beta &= -\frac{\log{2}}{6}+b_1^- e^{r\delta/L_-}+\cdots
\end{align}
where $\delta=1+\sqrt{7}$. The BPS equations imply
\begin{align}
    A_0^-&=\log \frac{L_-}{2} \nonumber\\ 
    A_2^-&=1\nonumber \\
    A_4^{-}&=-\frac{1}{2}\nonumber\\
    \alpha_1^-&=\frac{2}{3}\left( 4+\sqrt{7}\right)\zeta \nonumber\\
     b_1^-&=\frac{1}{18}\left( 3+2\sqrt{3}\right)\left(1+\sqrt{7} \right)\zeta
\end{align}
reproducing the results of \cite{Arav:2020asu}.

\section{On-shell action for holographic RG interfaces}\label{app:OS}
As detailed in appendix B\footnote{There the holographic renormalization of an asymptotically LS boundary was not considered, as the solutions discussed therein are generically holographic interfaces between mass deformed $\mathcal{N}=4$ theories.} of \cite{Arav:2020obl}, the full gravitational action relevant for the holographic RG interfaces discussed in the present work can be written as
\begin{equation}
S = S_{\mathrm{bulk}} + S_{\mathrm{GH}}^\pm + S_{\mathrm{CT}}^\pm + S_{\mathrm{finite}}^\pm
\end{equation}
where 
\begin{equation}
S^\pm_{\mathrm{GH}} = -\frac{1}{8\pi G_N}\int^\pm\dd ^4 x \sqrt{|\gamma|} K = \pm\frac{1}{2\pi G_N}\int^\pm\dd ^4 x \sqrt{|\gamma|} A'
\end{equation}
is the Gibbons-Hawking term (on each asymptotic AdS$_5$ boundary), while $S^\pm_{\mathrm{CT}}$ and $S_{\mathrm{finite}}^\pm$ are the divergent and finite boundary counterterms (respectively) required to renormalize the on-shell action in a scheme consistent with boundary supersymmetry. In this appendix, {\it we will compute field theory quantities for the field theory defined on a unit radius AdS$_4$}. Accordingly, here we find it convenient to regularize the solutions along constant $r=\mathfrak{r}_\pm$ surfaces.

Focusing first on the bulk contribution, we note that by employing the Wald formalism it is possible to write the on-shell bulk action as a sum of total derivatives. In particular, we obtain
\begin{align}\label{eq:Sbos}
S_{\mathrm{bulk}}^\mathrm{OS} & = \frac{1}{8\pi G_N}\int \dd^5 x \left[-\partial_r\left(\frac{A' \, r_{||}}{x^4}e^{4A} \right)+\partial_x\left(\frac{e^{2A} \, r_{||}}{x^3} \right) \right] \nonumber\\
& = -\frac{1}{8\pi G_N}\int \dd^4 x \sqrt{|\gamma|} A'\Big |^{\mathfrak{r}_+}_{\mathfrak{r}_-} - \frac{3}{8\pi G_N}\int \dd^5 x\, \frac{e^{2A} \, r_{||}}{x^4}.
\end{align}
Note that for BPS configurations, such as those studied in this work, on-shell the first of these integrals can be alternatively expressed in terms of the superpotential using
\begin{equation}
A' = \frac{1}{3}\mathrm{Re}\Big(e^{-i\xi+\mathcal{K}/2}\overline{\mathcal{W}} \Big).
\end{equation}
Interestingly, written in this way the on-shell bulk action in (\ref{eq:Sbos}) decomposes into the sum of two terms---the first a standard boundary term evaluated on the holographic boundary, and the second a bulk integral remainder. 

The latter is reminiscent of (\ref{eq:Amin}), and our first task is to understand its contribution to the on-shell action. Exploiting the $x$-independence of the $\mathfrak{r}_\pm$ cut-off surfaces in this case,  the bulk term can be written
\begin{align}
 -\frac{3}{8\pi G_N}\int \dd^5 x\, \frac{e^{2A} \, r_{||}}{x^4} & =  -\frac{3}{8\pi G_N}\cdot  \mathrm{vol}\left(\mathrm{AdS}_4 \right)\int^{\mathfrak{r}_+}_{\mathfrak{r}_-} \dd r\, e^{2A} \nonumber\\
 & =  -\frac{3}{8\pi G_N} \mathrm{vol}\left(\mathrm{AdS}_4 \right)\left[\mathcal{I}^+ + \int_{r_{0}^-}^{r_{0}^+} \dd r \,e^{2A} -  \mathcal{I}^-\right] \nonumber \\
 & = \frac{3}{8\pi G_N}\cdot \mathrm{vol}\left(\mathrm{AdS}_4 \right)\left[\mathcal{J}^+\left[r_0^+\right] - \int_{r_{0}^-}^{r_{0}^+} \dd r \,e^{2A} -\mathcal{J}^-\left[r_0^-\right]\right] \nonumber\\ & 
 \quad - \frac{3}{8\pi G_N}\cdot \mathrm{vol}\left(\mathrm{AdS}_4 \right)e^{2A_0^+}\left(2A_2^+\mathfrak{r}^+ + \frac{L_+}{2}
e^{2 \mathfrak{r}^+/L_+} \right)\nonumber\\ & 
 \quad + \frac{3}{8\pi G_N}\cdot \mathrm{vol}\left(\mathrm{AdS}_4 \right)e^{2A_0^-}\left(2A_2^- \mathfrak{r}^- - \frac{L_-}{2}
e^{-2 \mathfrak{r}^-/L_-} \right)
\end{align}
or, borrowing the notation of section \ref{sec:meth},
\begin{multline}
-\frac{3}{8\pi G_N}\int \dd^5 x\, \frac{e^{2A} \, r_{||}}{x^4} = \frac{3}{8\pi G_N}\mathrm{vol}\left(\mathrm{AdS}_4\right)\Bigg[\frac{2 G_N}{\pi} \mathcal{C}^I -e^{2A_0^+}\left(2A_2^+\mathfrak{r}^+ + \frac{L_+}{2}
e^{2 \mathfrak{r}^+/L_+} \right)\\+e^{2A_0^-}\left(2A_2^- \mathfrak{r}^- - \frac{L_-}{2}
e^{-2 \mathfrak{r}^-/L_-} \right) \Bigg].
\end{multline}

To proceed we will need the explicit form of the $S^\pm_{\mathrm{CT}}$ and $S_{\mathrm{finite}}^\pm$ for our gravitational theory. Addressing first the $\mathcal{N}=4$ SYM (`+') boundary, the relevant counterterm action from \cite{Arav:2020obl} in our conventions is 
\begin{multline}
S^+_{\mathrm{CT}} = \frac{1}{16\pi G_N}\int^+ \dd^4 x \sqrt{|\gamma|}\Bigg[-6+ \frac{1}{2}R-2\phi^2-4\left(1-\frac{1}{2r} \right)\left(6\beta^2+\alpha^2\right)\\ -r\left[\frac{1}{4}\left(R_{ab}R^{ab}-\frac{1}{3}R^2 \right)+\frac{1}{3}\phi^2 R-\frac{16}{3}\phi^4\right]+\ldots\Bigg]
\end{multline}
and 
\begin{multline}
S^+_\mathrm{finite} = \frac{1}{16\pi G_N}\int^+ \dd^4 x \sqrt{|\gamma|}\Bigg[-\frac{\delta_{R^2}}{4}\left(R_{ab}R^{ab}-\frac{1}{3}R^2 \right)-\frac{\delta_{\Delta R^2}}{4}\left(R_{ab}R^{ab}+\frac{1}{3}R^2 \right)\\-\frac{\delta_{R\phi^2(1)}}{3}\phi^2 R+\delta_{4(1)}\frac{16}{3}\phi^4-\delta_\alpha\frac{4}{r^2}\alpha^2-\delta_\beta \frac{24}{r^2}\beta^2 +\delta_{\tilde{\beta}}24\left(\frac{\beta}{r}+\frac{2}{3}\phi^2 \right)^2+\ldots\Bigg].
\end{multline}
In \cite{Arav:2020obl}, supersymmetry was used to partially fix a subset of the finite counterterms. In particular, we will take
\begin{equation}
\delta_{4(1)} = -\frac{1}{4}+2\delta_\beta.
\end{equation}

Note that the counterterm actions as written above are sufficient for computing the renormalized on-shell action for our family of holographic RG interfaces, but we have dropped terms that may be relevant for computing certain correlators, as indicated by the ellipses.

Similarly, the counterm action for the LS side of the interface includes the terms
\begin{equation}
S_{\mathrm{CT}}^- =  \frac{1}{16\pi G_N}\int^- \dd^4 x \sqrt{|\gamma|}\Bigg[-\frac{6}{L_-}+ \frac{L_-}{2}R+\frac{L_-^2}{4}r\left(R_{ab}R^{ab}-\frac{1}{3}R^2 \right)+ \ldots \Bigg]
\end{equation}
together with the finite counterterms 
\begin{multline}
S^-_\mathrm{finite} = \frac{L_-^3}{16\pi G_N}\int^- \dd^4 x \sqrt{|\gamma|}\Bigg[-\frac{\tilde{\delta}_{R^2}}{4}\left(R_{ab}R^{ab}-\frac{1}{3}R^2 \right)\\-\frac{\tilde{\delta}_{\Delta R^2}}{4}\left(R_{ab}R^{ab}+\frac{1}{3}R^2 \right)+\ldots\Bigg].
\end{multline}
As before, these are not the full counterterm actions required to renormalize the asymptotically LS boundary, but they will be sufficient for the solutions considered here.

After some calculation, we find that the renormalized on-shell action can be written
\begin{multline}\label{eq:Sos}
S^{\mathrm{OS}} = \frac{\mathrm{vol}\left(\mathrm{AdS}_4 \right)}{4\pi G_N}\Bigg[\frac{3 G_N}{\pi}\mathcal{C}^I +\phi_{(s)}\alpha_{(v)}+\frac{3}{16}\left(1+4\delta_{R^2} -28\delta_{\Delta R^2}\right)
-\left(\delta_{\alpha}+\delta_{R\phi^2(1)} \right)\phi_{(s)}^2 \\+ \frac{3}{16}L_-^3\left(1+4\tilde{\delta}_{R^2} -28\tilde{\delta}_{\Delta R^2} \right)\Bigg]
\end{multline}
where in our conventions $L_- = 3/2^{5/3}$. 

Interestingly, there is a choice of the coefficients governing the finite counterterms for which the on-shell action can be written in terms of the stress tensor one point function. To see this, we employ the results of \cite{Arav:2020obl} to write the renormalized stress tensor vacuum expectation value on the `$+$' side of the interface as
\begin{equation}
\langle \mathcal{T}^{\mathcal{N}=4}_{ab} \rangle\dd x^a \dd x^b = a_T^{\mathcal{N}=4} \,\dd s^2\left(\mathrm{AdS}_4 \right)
\end{equation}
where
\begin{equation}\label{eq:aTdef}
a_T^{\mathcal{N}=4} = \frac{1}{4}\Big(\mathcal{A}^{\mathcal{N}=4}-\phi_{(s)}\langle\mathcal{O}_\phi \rangle-2\alpha_{(s)}\langle\mathcal{O}_\alpha \rangle -2\beta_{(s)}\langle\mathcal{O}_\beta \rangle \Big).
\end{equation}
Here, $\mathcal{A}^{\mathcal{N}=4}$ is the conformal anomaly for the deformed $\mathcal{N}=4$ SYM theory on AdS$_4$. In particular, for our supersymmetric configurations it is given by
\begin{equation}
\mathcal{A}^{\mathcal{N}=4} = \frac{1}{8\pi G_N}\left[- \frac{1}{8}\left(R_{ab}R^{ab}-\frac{1}{3}R^2 \right)-\phi_{(s)}^2\left(1+\frac{1}{6}R \right)\right],
\end{equation}
where the curvature invariants are evaluated on the boundary AdS$_4$ metric. Note that (\ref{eq:aTdef}) is consistent with the trace Ward identity obeyed by the stress tensor. Explicit expressions for the renormalized scalar one point functions on AdS$_4$ are provided in appendix C.2.1 of \cite{Arav:2020obl}. In terms of the coefficients governing the asymptotic fall-offs near the $\mathcal{N}=4$ boundary, one finds
\begin{equation}\label{eq:aTNis4}
a_T^{\mathcal{N}=4} = \frac{1}{4\pi G_N}\left[\frac{3}{16}-\phi_{(s)}\alpha_{(v)}+\frac{1}{4}\left(1+4\delta_\alpha+2\delta_{R\phi^2(1)} \right)\phi_{(s)}^2 \right].
\end{equation}

The first term in square brackets is the expected Casimir energy for the theory on AdS$_4$, while the second and third encode the presence of the non-trivial deformation of the $\mathcal{N}=4$ theory by supersymmetric mass terms. 

An analogous computation for the `$-$' side of the interface gives
\begin{equation}\label{eq:aTLS}
a_T^{\mathrm{LS}} = \frac{L_-^3}{4\pi G_N}\left(\frac{3}{16} \right),
\end{equation}
again reproducing the Casimir energy for the LS vacuum on AdS$_4$.

\section{Interface entropy scaling for large deformations}\label{app:RGscaling}
In section \ref{sec:res}, the numerical computation of the interface contribution to the sphere entanglement entropy, $\mathcal{C}^I$, revealed a power law scaling for large mass deformations:
\begin{equation}\label{eq:appScale}
\mathcal{C}^I\left(|\phi_{(s)}|\to\infty\right) \sim  |\phi_{(s)}|^\gamma \qquad \mathrm{with} \qquad \gamma \approx 2.
\end{equation}
In this appendix we investigate this feature in further detail. 

In the large deformation limit, the bulk solutions asymptote to ``junction'' type solutions \cite{Arav:2020asu}. In particular, as $\phi_{(s)}\to -\infty$, the bulk profile approaches a domain wall between the AdS$_5^\mathrm{LS}$ vacuum and the holographic RG flow solution describing the Poincar\'e invariant RG flow between $\mathcal{N}=4$ SYM and the LS$^+$ SCFT. When $\phi_{(s)}\to +\infty$, on the other hand, the bulk profile approaches a domain wall solution between the dual of the Poincar\'e invariant RG flow between $\mathcal{N}=4$ SYM and the LS$^-$ SCFT, and a Janus-type solution between the AdS$_5^{\mathrm{LS}^\pm}$ vacua.

\begin{figure}[h]
\centering
\includegraphics[scale=0.8]{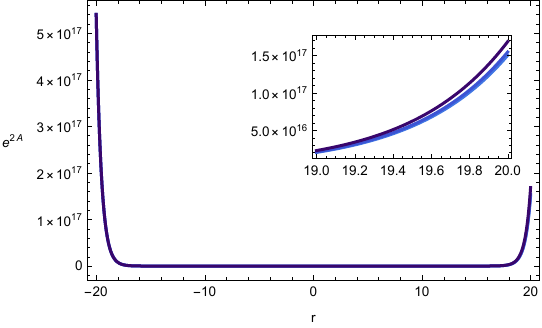}
\includegraphics[scale=0.8]{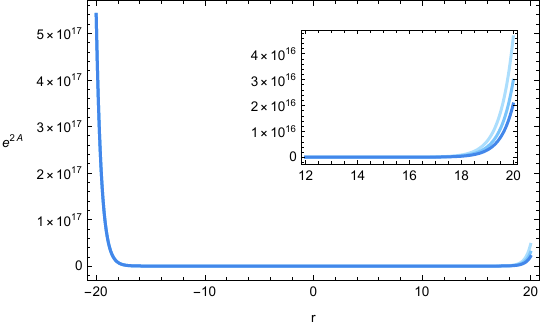}
\includegraphics[scale=0.8]{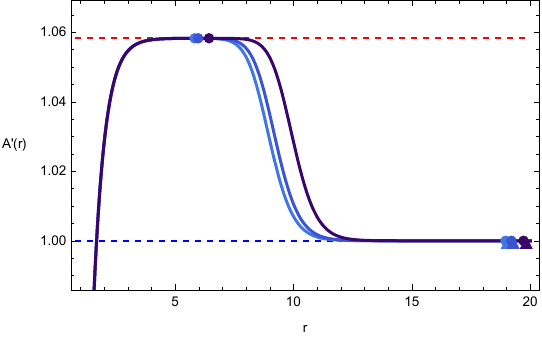}
\includegraphics[scale=0.8]{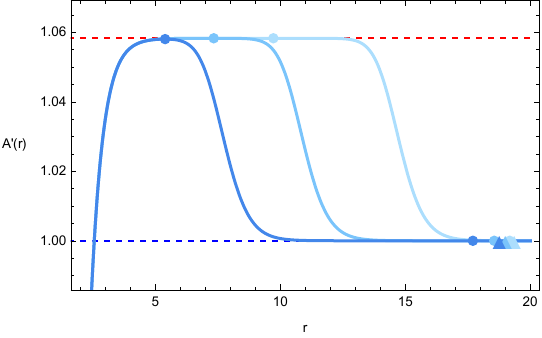}
\caption{The bulk radial profiles for $e^{2A}$ (top row) and $A'$ (bottom row) for several representative solutions with large negative $\phi_{(s)}$ (left column) and large positive $\phi_{(s)}$ (right column). The color coding is correlated with that of figure \ref{fig:ImzvsRez}. In the left column, the curves correspond to $\phi_{(s)}=\{-7759.4, -3553.2, -2728.9\}$, ordered from darkest to lightest. In the right column, the curves correspond to $\phi_{(s)}=\{288.6, 8154.3, 475367.7\}$, again ordered from darkest to lightest. In the plots along the bottom row, filled circles indicate the locations of $r_{\mathrm{IR}}$ and $r_{\mathrm{UV}}$, while filled triangles show the location $r_0^+$ where the numerical solution is glued to the analytic FG solution. The dashed lines are included to provide reference for the values $1/L_-$ (top) and $1/L_+$ (bottom).}\label{fig:RGscaleP}
\end{figure}

As the power law scaling of \ref{eq:appScale} occurs for either sign of large $\phi_{(s)}$, one may wonder if it is inherited from the common asymptotic ``Poincar\'e invariant RG flow'' region of the bulk geometry which appears in this limit. To explore this idea, we first note that as the limit is obtained, this RG flow region grows parametrically. This is illustrated in figure \ref{fig:RGscaleP}.

To identify the proper extent of the RG flow region, we mark two radial points $r_\mathrm{IR}$ and $r_\mathrm{UV}$ defined such that $r_\mathrm{UV}$ denotes the solution's transition from the AdS$_5^{\mathcal{N}=4}$ vacuum to the flow region, and $r_\mathrm{IR}$ denotes the transition from the flow region to the AdS$_5^\mathrm{LS}$ vacuum. Our operational definition of this transition is the points at which
\begin{equation}
A'(r_\mathrm{UV}) - \frac{1}{L_+} = 10^{-5} \qquad \mathrm{and} \qquad A'(r_\mathrm{IR}) - \frac{1}{L_-} = 10^{-5},
\end{equation}
as the solution is traversed in the direction of increasing $r$. Importantly, in each case these points satisfy the inequality
\begin{equation}
r_0^- < r_\mathrm{IR} < r_\mathrm{UV} < r_0^+
\end{equation}
which implies that the RG flow region as defined above is included in full in our computation of $\mathcal{C}^I$.

\begin{figure}[h]
\centering
\includegraphics[scale=0.85]{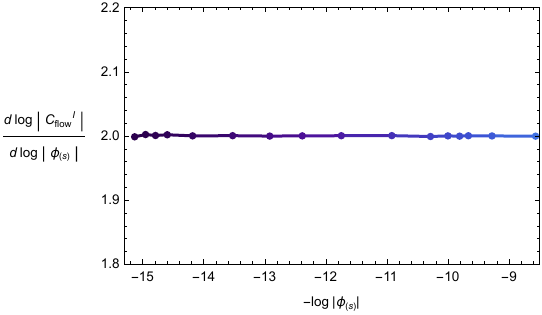}
\includegraphics[scale=0.85]{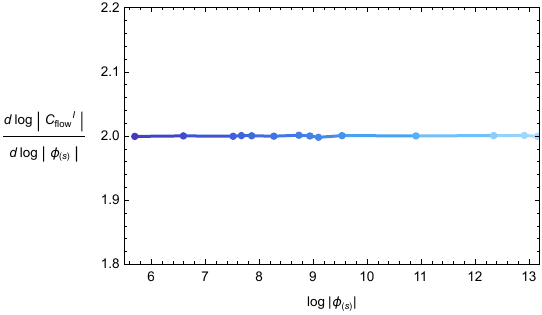}
\caption{The logarithmic derivative of $|\mathcal{C}^I_\mathrm{flow}|$ with respect to $\log|\phi_{(s)}|$ for large negative (left) and positive (right) deformations. The color coding is correlated with that of figure \ref{fig:ImzvsRez}.}\label{fig:CflowScale}
\end{figure}

We then continue by defining the bulk quantity $\mathcal{C}^I_\mathrm{flow}$, like
\begin{equation}
\mathcal{C}^I_\mathrm{flow} = \int_{r_\mathrm{IR}}^{r_\mathrm{UV}}\mathrm{d}r \,e^{2A} 
\end{equation}
as a proxy for the RG flow region's contribution to the interface entropy, as per (\ref{eq:NCI}). As illustrated in figure \ref{fig:RGscaleP}, the growth of the RG flow region is to a good approximation encoded in the $\phi_{(s)}$ dependence of the integration limits. Evaluating this integral for several large deformation magnitude solutions reveals that indeed
\begin{equation}
\mathcal{C}_{\mathrm{flow}}^I\left(|\phi_{(s)}|\to\infty\right) \sim  |\phi_{(s)}|^2.
\end{equation}
This behavior is shown clearly by the logarithmic derivative of $\mathcal{C}^I_\mathrm{flow}$ provided in figure \ref{fig:CflowScale}.

We note that this attribution of the asymptotic scaling observed in the full $\mathcal{C}^I$ to the parametric growth of the RG flow region sidesteps the numerical delicacy involved in forming the $r_0^\pm$ independent sum of (\ref{eq:NCI}). Thus, in addition to providing an explanation for the observed scaling, it offers evidence that this scaling is a robust feature of the RG interface solutions, rather than an artifact of a numerical failure to cancel the quadratic $\phi_{(s)}$ dependence in $\mathcal{J^+}$.

\section{Alternative schemes for RG interface entanglement entropy}\label{app:schemes}
Although we have argued for its physical sensibility, the subtraction scheme we've implemented to define $\mathcal{C}^I$ is clearly not unique. Once the quadratic and logarithmic divergences characteristic of the ambient theory have been subtracted, rescaling the UV cut-off leaves any finite term in the sphere entanglement entropy unchanged. Thus, when performing the ``background subtraction'', there is in principle a residual ambiguity in which part of the finite remainder one assigns to the ambient theory (to be subtracted), and which part is characteristic of the interface alone. It is thus natural to wonder what other schemes might be introduced to help quantify the entanglement properties of our RG interfaces, and how our results vary with the choice of subtraction scheme. Towards this end, in this appendix we investigate two other schemes for removing contributions from the ambient theory on either side of the interface.

\begin{figure}[h]
\centering
\includegraphics[scale=1.4]{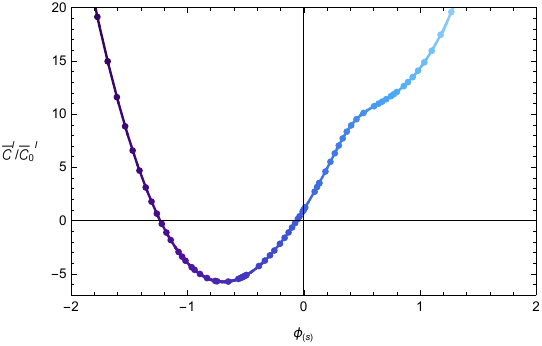}
\caption{The interface contribution to the sphere entanglement entropy for the RG interfaces, $\overline{\mathcal{C}}^I$, normalized by its value for the source free RG interface $\overline{\mathcal{C}}_0^I$. The color coding is correlated with that of figure \ref{fig:ImzvsRez}.}\label{fig:Vscheme}
\end{figure}

In the first, which we might refer to as a ``vacuum''  subtraction scheme, we remove the contribution to the entanglement entropy from the {\it undeformed} vacuum CFTs, as well as a logarithmically diverging remainder proportional to the source $\phi_{(s)}$:
\begin{equation}
\overline{\mathcal{C}}^I \equiv \frac{1}{4G_N}\left[\mathbb{A}_{\mathrm{min}}-\frac{1}{2}\left(\mathbb{A}_{\mathrm{vac}}^{\mathcal{N}=4}+\mathbb{A}_{\mathrm{vac}}^{LS}\right)-\left(\pi L_+^3-4\pi L_+ A_2^+e^{2A_0^+} \right)\log \frac{2R}{\epsilon} -s_1\left(\frac{R}{\epsilon} \right)\right]
\end{equation}
where we identify from (\ref{eq:vacSEE}) the contribution
\begin{equation}
\mathbb{A}_{\mathrm{vac}}^{\mathcal{N}=4/\mathrm{LS}} = 2\pi L^3_\pm\left(\frac{R^2}{\epsilon^2}-\log\frac{2R}{\epsilon}-\frac{1}{2} \right).
\end{equation}
This quantity, normalized by its value at the source-free RG interface, is shown in figure \ref{fig:Vscheme}. 

An alternative scheme, the ``improved vacuum'' scheme, we define by noting that the divergence structure of the vacuum entanglement entropy encodes (scheme dependent) vacuum contributions in both the quadratic and logarithmic terms. In the interface solutions studied here, the coefficient of the log divergence is given in terms of the solution's asymptotic data as
\begin{equation}
c_{\mathrm{log}} = -4\pi L_\pm A_2^\pm e^{2A_0^\pm}.
\end{equation}

On the $\mathcal{N}=4$ side, the deformation parameter $\phi_{(s)}$ appears in this quantity via its dependence on $A_2^+$. Noting that the combination $c_{\mathrm{log}} $ also appears in the finite part of $\mathbb{A}_{\mathrm{min}}$, this scheme identifies that contribution as characteristic of the vacuum and subtracts it as well. In other words, in the ``improved vacuum'' scheme the interface contribution to the entanglement entropy is given by
\begin{multline}
\hat{\mathcal{C}}^I=\frac{1}{4G_N}\Bigg[\mathbb{A}_{\mathrm{min}}-\pi\left(L_+^3 + L_-^3 \right)\left(\frac{R}{\epsilon} \right)^2+4\pi\left(L_+ A_2^+ e^{2A_0^+}+L_- A_2^- e^{2A_0^-} \right)\ln\left(\frac{2R}{\epsilon} \right)\\-s_1\left(\frac{R}{\epsilon} \right)-\frac{\pi}{2}\left(L_+^3 + L_-^3 \right) +4\pi \left(L_+ A_2^+ e^{2A_0^+}+L_- A_2^- e^{2A_0^-}\right) \Bigg].
\end{multline}

\begin{figure}[h]
\centering
\includegraphics[scale=1.4]{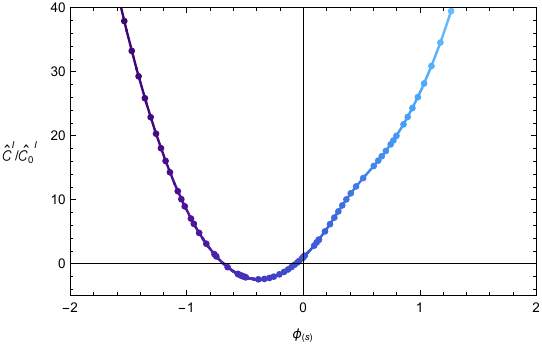}
\caption{The interface contribution to the sphere entanglement entropy for the RG interfaces, $\hat{\mathcal{C}}^I$, normalized by its value for the source free RG interface $\hat{\mathcal{C}}_0^I$. The color coding is correlated with that of figure \ref{fig:ImzvsRez}.}\label{fig:imVscheme}
\end{figure}

 This quantity is plotted as a function of $\phi_{(s)}$ in figure \ref{fig:imVscheme}.
While differing in the details, the broad stroke features of these interface entropies are shared---the existence of local extrema, multiple zeroes, and quadratic divergence at large values of $|\phi_{(s)}|$.   We have investigated a number of additional schemes, and found these features to be fairly robust.
 
\end{appendix}

\bibliography{defectsRGI}{}
\bibliographystyle{utphys}

\end{document}